%% file: convDictTomo.tex
\documentclass{article}

\include{package}
\include{defs}

\begin{document}

\title{Convolutional Dictionary Regularizers for Tomographic Inversion}

\name{S.~V.~Venkatakrishnan$^{\star}$
  \thanks{\scriptsize{This manuscript has been authored by UT-Battelle, LLC, under Contract No. DE-AC05-00OR22725 with the U.S. Department of Energy. 
The United States Government and the publisher, by accepting the article for publication, acknowledges that the United States Government retains a non-exclusive, paid-up, irrevocable, world-wide license to publish or reproduce the published form of this manuscript, or allow others to do so, for United States Government purposes.
DOE will provide public access to these results of federally sponsored research in accordance with the DOE Public Access Plan (http://energy.gov/downloads/doe-public-access-plan).
S.V. Venkatakrishnan and B.Wohlberg were supported via the Laboratory Directed Research and Development program
at Oak Ridge National Lab and Los Alamos National Lab respectively. 
}}
  \qquad Brendt Wohlberg $^{\dagger}$}
\address{$^{\star}$ Imaging, Signals and Machine Learning Group, Oak Ridge National Lab, Oak Ridge, TN 37831 \\
$^{\dagger}$ Theoretical Division, Los Alamos National Laboratory, Los Alamos, NM 87545}

\maketitle

\input{front}

\input{convDictTomoBody}

\bibliographystyle{IEEEbib}
\footnotesize
\bibliography{convDictTomo}

\end{document}

%% file: package.tex
\usepackage{mathtools}
\usepackage{mathdesign}
\usepackage{spconf}
\usepackage{array}
\usepackage{mdwmath}
\usepackage{mdwtab}
\usepackage{fixltx2e}
\usepackage{url}
\usepackage{color}
\usepackage[]{graphicx}
\usepackage{epsfig}
\usepackage{bbm}
\usepackage{enumerate}
\usepackage{amsfonts}
\usepackage{amsmath}
\usepackage{amssymb}
\usepackage{cite}
\usepackage{algorithm}
\usepackage{algpseudocode}
\usepackage{eqparbox}
\usepackage{comment}
\usepackage{breqn}
\usepackage{dblfloatfix}

%% file: defs.tex
\providecommand{\norm}[1]{\lVert#1\rVert}

\newcommand{\remark}[1]{%
{\textcolor{cyan}{[\small #1]}}%
}%
\newcommand{\rmksv}[1]{%
{\textcolor{red}{[SV: \small #1]}}%
}
\newcommand{\rmkbw}[1]{%
{\textcolor{blue}{[BW: \small #1]}}%
}
\renewcommand{\remark}[1]{}  
\renewcommand{\rmksv}[1]{}
\renewcommand{\rmkbw}[1]{}

%% file: front.tex
\begin{abstract}
  There has been a growing interest in the use of 
  data-driven regularizers to solve inverse problems associated with
  computational imaging systems.
  The convolutional sparse representation model 
  has recently gained attention,
  driven by the development of fast algorithms for solving
  the dictionary learning and sparse coding problems for sufficiently
  large images and data sets.
  Nevertheless, this model has seen very limited application to tomographic reconstruction problems.
  In this paper, we present a model-based tomographic reconstruction 
  algorithm using a learnt convolutional dictionary as a regularizer. 
  The key contribution is the use of a data-dependent
  weighting scheme for the $l_1$ regularization to construct an effective denoising method that is integrated into the inversion using
  the Plug-and-Play reconstruction framework.
  Using simulated data sets we demonstrate 
  that our approach can improve performance
  over traditional regularizers
  based on a Markov random field model and a patch-based sparse representation model 
  for sparse and limited-view tomographic data sets. 
\end{abstract}

%% file: convDictTomoBody.tex
\section{Introduction}

Model-based reconstruction algorithms have enabled
dramatic improvements in the performance
of tomographic reconstructions compared to traditional
approaches, especially for sparse, limited-view and noisy data sets~\cite{VenkatHAADF13}.
These methods solve the tomographic reconstruction by minimizing
a cost function that balances a data-fidelity term 
and a regularization term that promotes certain desirable
properties of the reconstruction itself. 
While model-based methods have helped to improve 
the performance of tomographic imaging systems, 
the potential to further improve the quality by using different  
regularizers is still being explored. 

Several regularizers have been proposed
to improve the quality of model-based
tomographic reconstructions.
These include the edge-preserving total-variation model~\cite{SaBo92}, 
the non-local self-similarity model~\cite{zhang2014nonlocal}, and 
those that constrain the solution to a sparse combination of elements from an over-complete dictionary based on wavelets or other transforms.
Data-driven regularizers that learn a model from an off-line database have also been applied to tomographic inversion~\cite{xu2012low, zhang2016gaussian, zheng2016low}.
%
Among data-driven regularizers, patch based dictionary
models~\cite{xu2012low, zheng2016low, luo20162} have been widely developed 
for tomography, with promising performance. 
However, the patch is a local model and can result in redundant dictionary elements that are merely translated versions of each other.
As a result there has been a revival of interest in the use of
shift-invariant~\cite{lewicki1999coding} models for images, also
called convolutional sparse representation (CSR) models~\cite{zeiler2010deconvolutional, wohlberg2016efficient}.
Recent work on efficient algorithms for convolutional sparse coding (CSC)~\cite{bristow2013fast,wohlberg2014efficient} and the corresponding 
convolutional dictionary learning (CDL) problem~\cite{garcia2018convolutional, papyan2017convolutional, chun2018convolutional,degraux2017online, liu2018first}
have allowed for the use of CSR as regularizers for a variety of inverse problems~\cite{gu2015convolutional, liu2016image, serrano2016convolutional, zhang2017convolutional}.

Existing approaches to exploiting the CSR model for tomography and related problems~\cite{quan2016compressed, skau2018tomographic} have integrated the inversion into a CDL problem, simultaneously learning the dictionary and the reconstruction as part of the optimization algorithm. This has the advantage of not requiring any ground-truth reconstructions for use as training data for learning a dictionary, but the integration into the dictionary learning process imposes some practical constraints on how the convolutional representation is exploited, and it is reasonable to expect that highly under-determined problems may benefit from a pre-trained dictionary if suitable training data are available. 
In this paper, we propose a tomographic inversion algorithm based on the CSR model, using a dictionary learnt from an external database~\cite{de2018tomobank}. Instead of directly integrating the inversion into a CSC problem, which would retain some of the difficulties that have to be addressed in the CDL-based approach discussed above, we use the Plug-and-Play (PnP) framework~\cite{VenkatPlugPlay13, sreehari-2016-plug} to couple the tomographic inversion with a CSC model that plays the role of a Gaussian white-noise denoiser.

\section{CSR for Image Denoising}
\label{sec:cd_denoise}

One of the simplest computational imaging problem is that of
recovering an image corrupted by additive white Gaussian noise.
As a result, this simple inverse problem serves as a convenient test-bed for the
development of new regularizers before extending them to other inverse problems. This extension is greatly simplified by the PnP method, which provides a simple method to integrate complex models expressed via denoising algorithms into the model-based inversion framework~\cite{VenkatPlugPlay13, sreehari-2016-plug}. Here we summarize different approaches to using CSR for solving the white-noise denoising problem. 

A convolutional dictionary, $\tilde{d}$, 
is typically learnt from a set of $K$ images by minimizing 
\begin{eqnarray*}
  c(d,\alpha)=\frac{1}{2}\displaystyle \sum_{k=1}^{K} \Big\|y_{k,h}-\displaystyle \sum_{m=1}^{M} d_m*\alpha_{k,m}\Big\|_{2}^{2} + \\
  \lambda \displaystyle \sum_{k=1}^{K} \displaystyle \sum_{m=1}^{M} \norm{\alpha_{k,m}}_{1}
\end{eqnarray*}
such that $\norm{d_m}_{2}=1$ $\forall m \in \{1,...M\}$, where $y_{k,h}$ is the high-pass component of the $k^{\text{th}}$
image, $d_m$ is the $m^{\text{th}}$ dictionary element, $\alpha_{k,m}$ is the
coefficient map corresponding to image $k$ and dictionary element $m$,
$\lambda$ is a regularization parameter that controls the sparsity of the coefficient maps, and $*$ is the convolution operator.
The dictionaries are learned from high-pass filtered training images rather than the original training images due to the difficulty in representing the low-pass components via convolutional sparse representations~\cite[Sec. 3]{wohlberg2016convolutional2}.
The high-pass component is typically set as
$y_{k,h}=y_{k}-(I+\lambda_{\text{LPF}}G^{t}G)^{-1}y_k$, where $G$ is a
2-D finite difference operator and $\lambda_{\text{LPF}}$ controls the strength of the filter~\cite{wohlberg2017sporco}. 
There are several algorithms for efficiently solving the CDL problem
\cite{papyan2017convolutional, degraux2017online, liu2018first, chun2018convolutional, garcia2018convolutional}. 

We consider three different variants of the CSC problem for white-noise denoising. The first approach, henceforth referred to as CSC-I, corresponds to the standard CSC problem, based on minimizing the function
\begin{eqnarray}
  c_{1}(\alpha) = \frac{1}{2}\Big\|y_{n,h}-\displaystyle\sum_{m=1}^{M} \tilde{d}_m*\alpha_m\Big\|_{2}^{2} + \lambda \displaystyle\sum_{m=1}^{M} \norm{\alpha_m}_{1}
\label{eq:cdn1}
\end{eqnarray}
where $y_{n,h}$ is the high-pass component of the noisy data $y_{n}$, computed in the same way as for dictionary learning.
The final reconstruction is obtained as $\displaystyle\sum_{m=1}^{M} \tilde{d}_m*\tilde{\alpha}_m + y_{n,l}$
where $y_{n,l}=y_{n}-y_{n,h}$ is the low-pass component of the noisy input image.

Since the standard CSC problem does not provide competitive performance in Gaussian white-noise denoising problems, we introduce a simple $\ell_1$ weighting scheme that has been found to significantly improve performance in this application, making it competitive with more well-established patch-based sparse representation methods~\cite{wohlberg-2017-convolutional3}. This variant,  henceforth referred to as CSC-II, can be expressed as the minimization of
\begin{eqnarray}
  c_{2}(\alpha)=\frac{1}{2}\Big\|y_{n,h}-\displaystyle\sum_{m=1}^{M} \tilde{d}_m*\alpha_m\Big\|_{2}^{2} + \lambda \displaystyle\sum_{m=1}^{M} \norm{w_{m} \odot \alpha_m}_{1}
\label{eq:cdn2}
\end{eqnarray}
where $\odot$ represents point-wise multiplication, and $w_m$ are weights that are set as
\begin{equation}
    w_m = 1 / (\tilde{D}_{m}^{T}y_{n,h})^{2} \;,
\end{equation}
where $\tilde{D}$ is the matrix version of the convolutional dictionary and $1/$ denotes point-wise division~\cite{wohlberg-2017-convolutional3}.

While the need for pre-processing of the input images is not problematic when solving a simple denoising problem, it greatly complicates direct integration of the CSC model with a more complex inverse problem since the input images and reconstructions are in different spaces.
The final variant we consider avoids the need for high-pass filtering pre-processing of the input images by jointly estimating low-pass and high-pass components, using an additional regularization term that penalises the gradient of the low-pass component~\cite{wohlberg2016convolutional2}. This problem , henceforth referred to as CSC-III, can be expressed as the minimization of 
\begin{eqnarray}
  c_{3}(\alpha)=\frac{1}{2}\Big\|y_n-\displaystyle\sum_{m=1}^{M} \tilde{d}_m*\alpha_m - \alpha_{M+1}\Big\|_{2}^{2} + \nonumber \\
  \lambda \displaystyle\sum_{m=1}^{M} \norm{\alpha_m}_{1} + \frac{\mu}{2} \norm{G\alpha_{M+1}}_{2}^{2} \;,
\label{eq:cdn3}
\end{eqnarray}
where $\lambda$ and $\mu$ are algorithm parameters.
The final reconstruction is obtained as $\displaystyle\sum_{m=1}^{M} \tilde{d}_m*\tilde{\alpha}_m + \tilde{\alpha}_{M+1}$.
While the use of the approach of CSC-III or variants thereof is necessary when directly integrating the CSR model with tomographic inversion~\cite{skau2018tomographic}, the decoupling provided by the PnP approach makes it optional rather than essential.

\section{CSR for Tomography}
\label{sec:cd_pnp}

To leverage the idea underlying the weighted convolutional sparse coding based denoising of \eqref{eq:cdn2} for tomography, we utilize the Plug-and-Play priors framework \cite{VenkatPlugPlay13}. 
The framework was originally inspired by solving a regularized inversion using the idea of variable splitting followed by use of the alternating direction method of multipliers \cite{boyd2011distributed} that results in iteratively solving two sub-problems corresponding to an inversion step followed by a denoising step. 
Furthermore, it was empirically observed that the algorithm converges to a fixed point even if arbitrary denoisers are used in the iterative framework. 
Specifically, the PnP reconstruction \cite{VenkatPlugPlay13} is obtained by iterating over the steps
\begin{eqnarray*}
  \tilde{x} &=& \hat{v}-u \\
   \hat{x} &\leftarrow& F(y,\tilde{x};\beta) \\
  \tilde{v} &=& \hat{x}+u \\
   \hat{v} &\leftarrow& H(\tilde{v};\lambda) \\
    u &=& u + (\hat{x}-\hat{v}) \;,
\end{eqnarray*}
where $F$ corresponds to an optimization problem corresponding to the forward model,
$H$ is a denoising algorithm, and $\beta$ and $\lambda$ are
algorithm parameters.
In particular, for conventional tomography problems, $F$ is given by
\begin{eqnarray}
  F(y,\tilde{x};\beta) \leftarrow \arg\min_{x} \left\{\frac{1}{2}( \norm{y-Ax}_{W}^2 + \beta  \norm{x-\tilde{x}}_2^{2})\right\}
  \label{eq:f_prob}
\end{eqnarray}
where $y$ is a vector of tomographic projection measurements, $A$ is the projection matrix, and $W$ is a diagonal matrix of weights corresponding to the inverse variance of the noise.
While solving \eqref{eq:f_prob} is expensive, in practice we partially solve it using a few iterations of an iterative algorithm.

The denoiser corresponding to CSC-II (Eq. \eqref{eq:cdn2}), $H(\tilde{v};\lambda)$,
is given by
\begin{flalign*}
\tilde{v}_{l} &\leftarrow (I+\lambda_{\text{\text{LPF}}}G^{t}G)^{-1}\tilde{v} \\
\tilde{v}_{h} &\leftarrow \tilde{v} - \tilde{v}_{l} \\
w_{m} &\leftarrow 1/(\tilde{D}_{m}^{T}\tilde{v}_{h})^2 \text{     }\forall m \in {0,...,M-1}\\
\tilde{\alpha} &\leftarrow \arg\min_{\alpha} \Big\{ \frac{1}{2}\Big\|\tilde{v}_{h} - \displaystyle\sum_{m=0}^{M-1} \alpha_m*\tilde{d}_m\Big\|_2^{2} \\
&+\lambda\displaystyle\sum_{m=0}^{M-1} \norm{w_m \odot \alpha_m}_{1} \Big\} \\
\hat{v} &\leftarrow \displaystyle\sum_{m=0}^{M-1} \tilde{d}_m*\tilde{\alpha}_{m} + \tilde{v}_{l} \;.
\end{flalign*}
We use a fixed dictionary $\tilde{d}$ that is learnt from an offline database for the 
proposed algorithm.
Notice that even though these sequence of steps do not correspond to solving an optimization problem, the PnP method allows for CSC-II (and by extension the CSR model) to be used for tomographic inversion.  

\section{Results}
\label{sec:results}
\begin{figure}[!htbp]
    \includegraphics[scale=0.28,trim=2.25cm 3cm 1.5cm 2cm,clip]{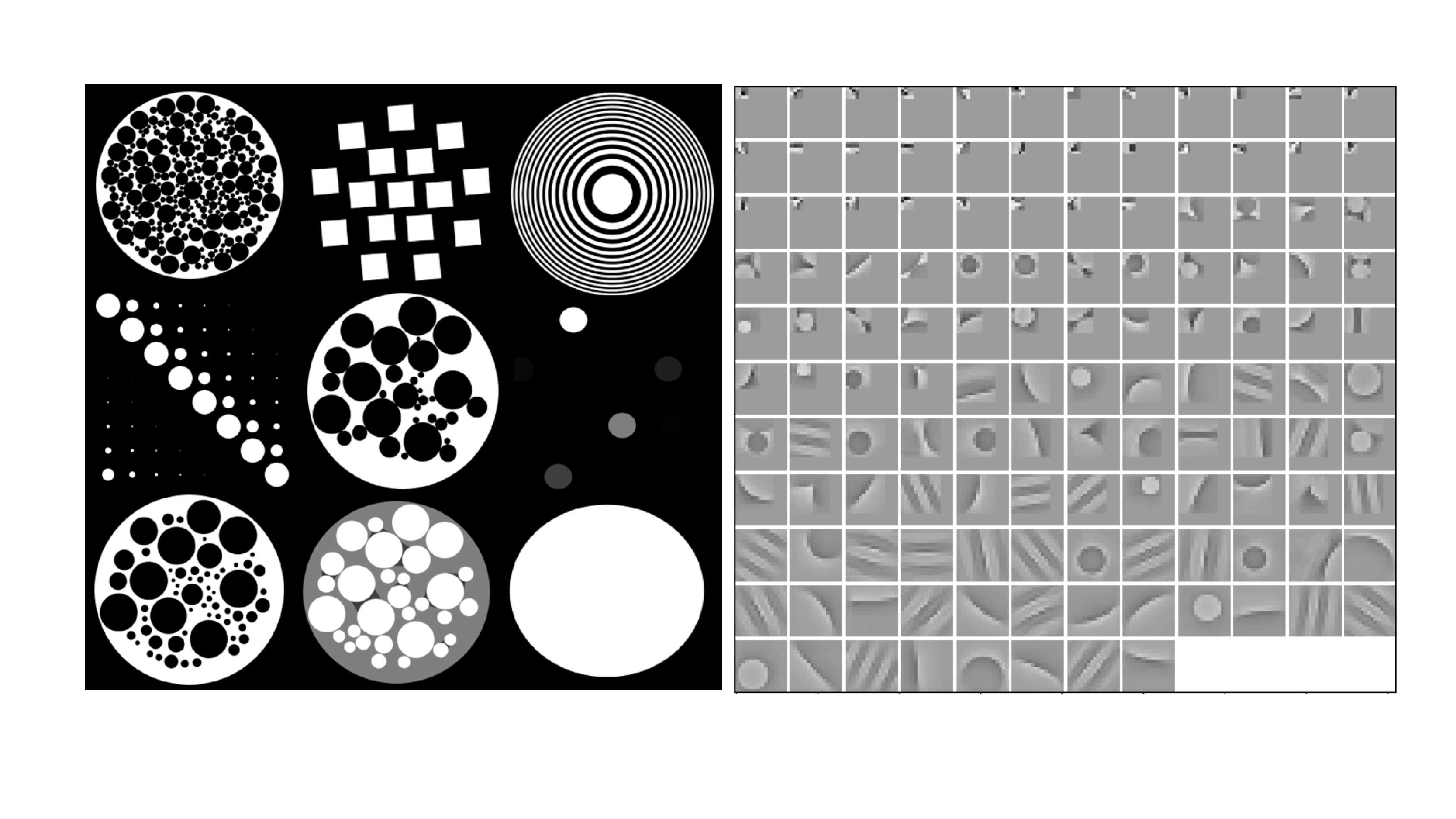} 
\caption{\label{fig:data_bank_dict} Training images and corresponding learnt convolutional dictionary.}
\end{figure}
\begin{table}[!h]
  \caption{Comparison of the denoising performance of the three convolutional sparse coding based denoisers in Section \ref{sec:cd_denoise} on measurements with different noise levels. All values represent PSNR values in units of dB.}
\label{tab:denoisePSNR}
\begin{center}
  \begin{tabular}{| c | r | r | r |}
    \hline
    Input PSNR & CSC-I & CSC-II & CSC-III \\
    \hline
    $34.15$  & $35.01$ & $\textbf{37.32}$ & $34.74$ \\
    \hline
    $24.60$   & $27.73$ & $\textbf{31.00}$ & $27.60$ \\
    \hline
    $20.17$  & $25.64$ & $\textbf{28.73}$ & $25.43$ \\
    \hline
    $14.50$  & $23.50$ & $\textbf{24.51}$ & $22.75$ \\
    \hline
  \end{tabular}
\end{center}
\end{table}
\begin{figure}[!h]
\begin{center}
\includegraphics[scale=0.5,trim=2.7cm 7.1cm 2cm 1.5cm,clip]{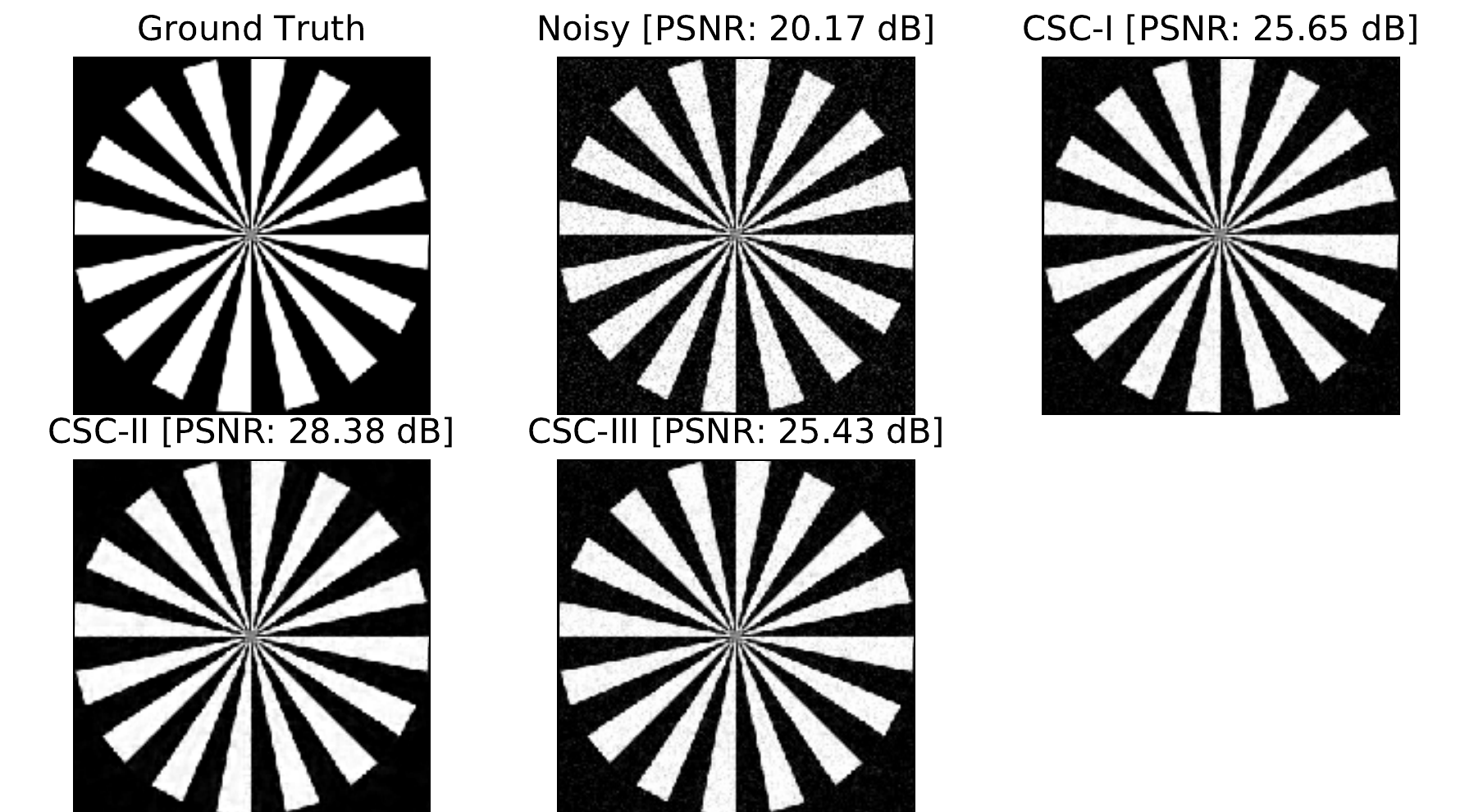} 
\end{center}
\vspace{-0.15in}
\caption{\label{fig:denoise} Illustration of denoising performance using the three different convolutional sparse coding based methods in Section \ref{sec:cd_denoise}.
Notice the weighted scheme (CSC-II) produces a high quality reconstruction and has a similar computational complexity to CSC-I.}
\end{figure}
\begin{table*}[!t]
  \caption{Comparison of the PSNR 
    of tomographic reconstructions with respect to the original
    phantom for different levels of noise.}
\label{tab:sparsePSNR}
\begin{center}
  \begin{tabular}{| c | c | c | c |}
    \hline
    PSNR & 26 dB & 20 dB & 14 dB \\
    Views & MRF | PSC | CSC & MRF | PSC | CSC & MRF | PSC | CSC \\
    \hline
    256 & \textbf{25.07} | 21.41 | 24.84 & 23.23 | 20.68 | \textbf{23.87} & 22.17 | 20.07 | \textbf{22.57} \\
    \hline
    128 & 22.18 | 20.33 | \textbf{22.99} & 21.07 | 19.80 | \textbf{21.96} & 20.06 | 19.06 | \textbf{20.68} \\
    \hline 
    64 & 18.71 | 18.09 | \textbf{20.79} & 18.10 | 17.69 | \textbf{19.42} & 17.22 | 17.05 | \textbf{17.87} \\
    \hline
  \end{tabular}
\end{center}
\end{table*}
\begin{table*}[!h]
  \caption{Comparison of the PSNR of the limited-angle tomographic reconstruction with 
    respect to the original phantom for various levels of noise.}
\label{tab:limitedPSNR}
\begin{center}
  \begin{tabular}{| c | c | c | c |}
    \hline
    PSNR & 26 dB & 20 dB & 14 dB \\
    Views & MRF | PSC | CSC & MRF | PSC | CSC & MRF | PSC | CSC \\
    \hline
    70 & 28.43 | 28.12 | \textbf{28.60} & 26.64 | \textbf{27.82} | 27.41 & 25.24 | \textbf{27.70} | 26.21 \\
    \hline
  \end{tabular}
\end{center}
\end{table*}
In order to test the proposed algorithm, we use
phantoms from the tomo-bank database \cite{de2018tomobank}.
Fig.~\ref{fig:data_bank_dict} shows the nine $256 \times 256$ images from the database that we used to
train a multi-scale convolutional dictionary with 128 elements of sizes
$2 \times 2$, $4 \times 4$, $8 \times 8$ and $16 \times16$.
We use the SPORCO package \cite{wohlberg2016sporco, wohlberg2017sporco} for implementing the CDL and CSC algorithms, using a value of $\lambda_{\text{LPF}}=7$ for pre-processing the images.
First, we test the different denoising strategies discussed in Section \ref{sec:cd_denoise}. 
Table.~\ref{tab:denoisePSNR} and Fig.~\ref{fig:denoise} shows the results of different denoising strategies using an image from the database that is not in the training set. 
Notice that the weighted scheme (CSC-II) offers superior performance
across different noise levels making it a useful method for using CSR as a regularizer for inverse problems.
We believe that this is because the weighting scheme encourages the use of the 
dictionary elements that are strongly correlated with the 
underlying image features while discouraging fits to the 
noise.
More importantly, CSC-II has the simplicity of CSC-I and offers computational savings compared to CSC-III along with better performance. \rmkbw{CSC-III is really a lot more expensive? I expect quite a minor increase in computational cost.}
\rmksv{I thought the extra optimization variable should add more memory as well computation, though I think I will remove the word significant as I am not sure whether I have a solid backing for it other than the code just running slower.}
\rmkbw{The CSC with TV problem is a lot slower due to the need for additional variables, but CSC with l2 of gradient (CSC-III) doesn't need any additional variables. I seem to recall that it's only about 10\% or so slower than the standard version, but perhaps I'm not remembering correctly. How much slower is it in your estimation?}\rmksv{For a 256X256 image all run on GPU, CSC-II takes ~3 sec, CSC-III takes 4 seconds. I did not compare per-iteration time, but the speed to convergence seems faster.}

\begin{figure}[!h]
  \includegraphics[scale=0.28,trim=2.5cm 0.7cm 1cm 0cm,clip]{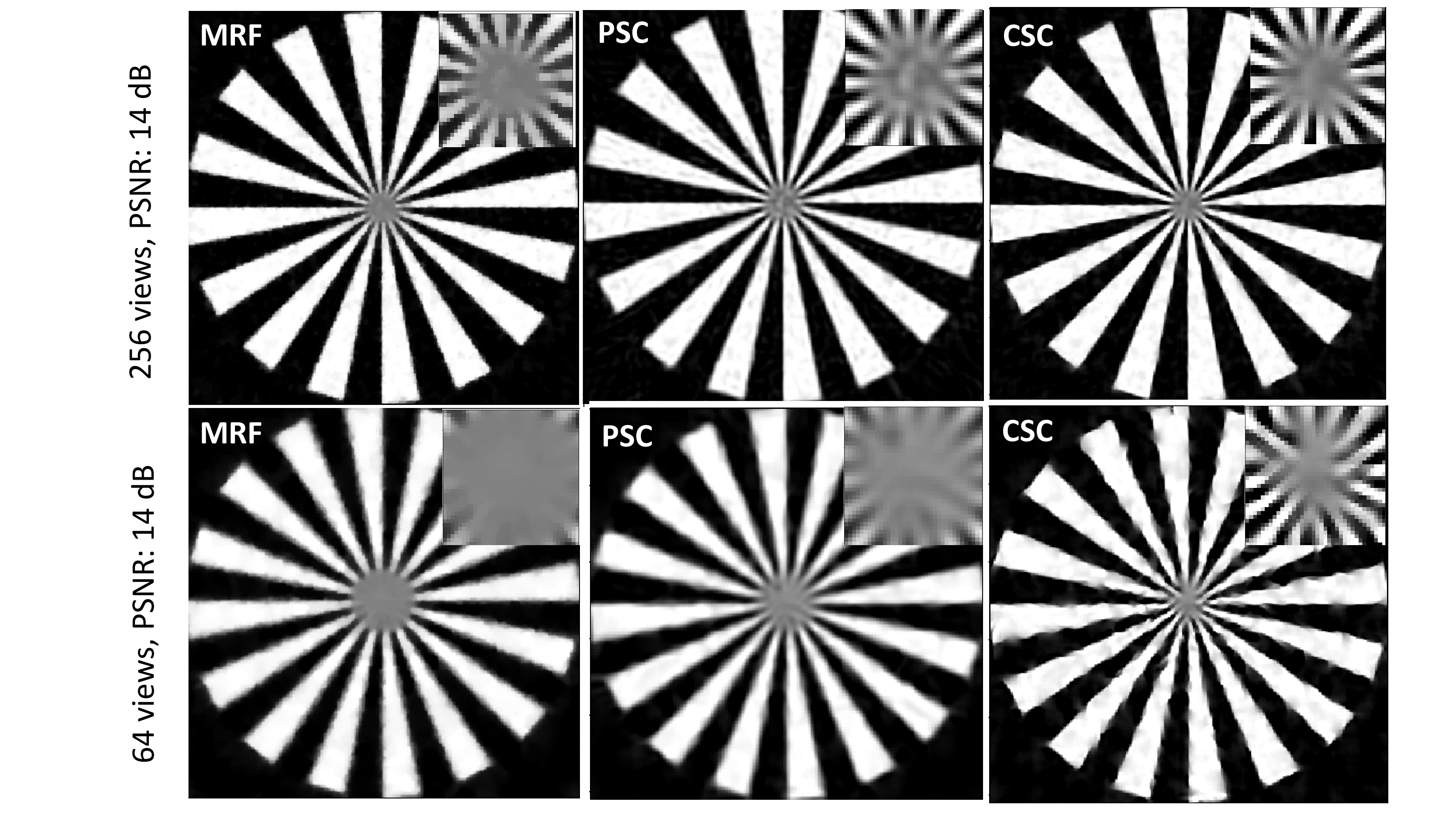}
\caption{\label{fig:sparse_view} Tomographic reconstruction using different algorithms for
  sparse-view and noisy data corresponding to the phantom in Fig.~\ref{fig:denoise}. The inset
  is a zoomed in section from the center of the reconstruction.}
\end{figure}
\begin{figure}[!h]
  \includegraphics[scale=0.28,trim=1cm 3.7cm 2cm 3.8cm,clip]{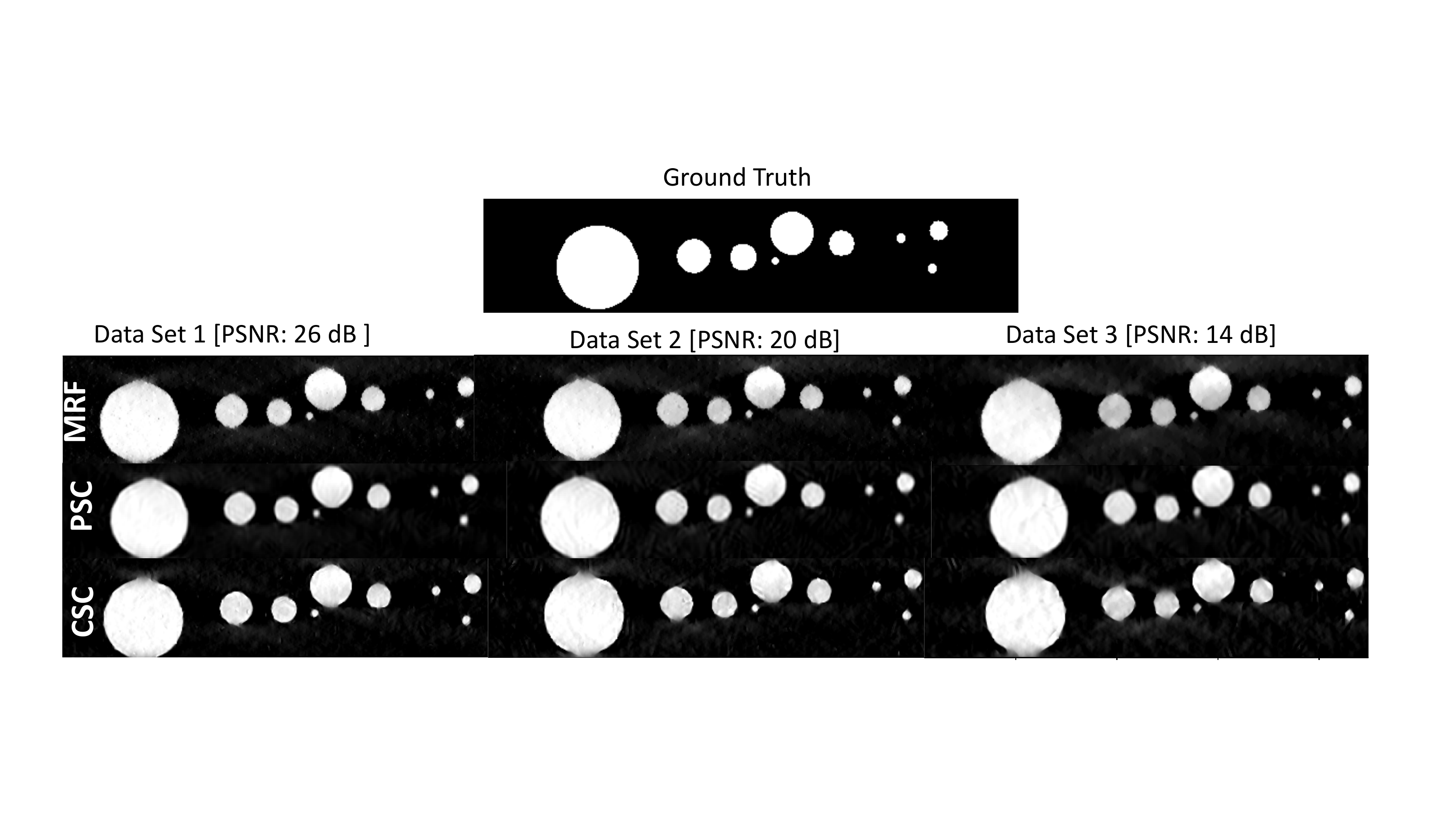}
\caption{\label{fig:limited_view} Ground-truth image and limited-angle tomographic reconstructions using different algorithms.}
\end{figure}
Next, we test the performance of the proposed algorithm on tomographic data sets.
We compare the proposed algorithm to a model-based algorithm using the 
edge-preserving Markov-random field (MRF) regularizer \cite{VenkatHAADF13} 
(with $p=1.2$) and a patch-based dictionary learning regularizer with a dictionary of $128$ elements of size $16 \times 16$ learnt from the database.
The patch model is used for tomography using the PnP framework with appropriate
averaging carried out to perform the image denoising \cite{wohlberg2017sporco}
and referred to in the results as patch-based sparse coding (PSC).
The $W$ matrix for the tomographic inversion is set to the identity matrix since the measurements are corrupted by i.i.d Gaussian noise. 
In all cases the parameters are chosen for the highest PSNR in the reconstruction.
The tomographic data was generated by projecting the phantom (obtained from the tomo-bank database \cite{de2018tomobank}) in Fig.~\ref{fig:denoise}
at $64$, $128$ and $256$ views with three different levels of
Gaussian noise corresponding to a PSNR of $26$ dB, $20$ dB and $14$ dB in the projection domain. 
The tomographic projection and back-projection was implemented using ASTRA tool-box \cite{BleichrodtAstra16}.
For the $F$ sub-problem in the PnP framework, we use
$25$ iterations of the optimized-gradient method \cite{KimOGM15} 
and for the $H$ sub-problem we set the maximum number of iterations to $25$. The total number of outer-iterations is set to $300$.
Fig.~\ref{fig:sparse_view} and Table~\ref{tab:sparsePSNR}
show the reconstructed results from different scenarios.
Notice that at the low-noise and large-views case 
the MRF model outperforms both the PSC and the CSC models.
However, for the sparse view and noisy data the
CSC model significantly outperforms the MRF and the PSC model.
Finally, we test the proposed algorithm on a limited
angle noisy data set, motivated by applications such as
electron tomography \cite{VenkatHAADF13}. 
The data was generated by projecting the phantom shown in Fig.~\ref{fig:limited_view}
at $70$ angles between $20^{\circ}$ and $160^{\circ}$ at the
three noise levels as in the first case. 
Fig.~\ref{fig:limited_view} and Table~\ref{tab:limitedPSNR}
highlight that the proposed algorithm
performs better than the MRF model in each case and is competitive 
to the PSC model for the high-noise cases. 

\section{Conclusion}
\label{sec:Concl}
We have presented an algorithm for using the convolutional sparse representation model as a regularizer for solving tomographic inverse problems.
To overcome the potentially poor performance of conventional CSR based approaches we use a data-adaptive weighting based denoising with the plug-and-play framework for the tomographic inversion.
This weighting is simple yet vital to boosting
the performance of the CSR model compared to the conventional edge-preserving model and the 
patch-based sparse representation model.
\rmkbw{References need a bit of cleaning up (I have already done quite a bit):
\begin{itemize}
    \item Inconsistent author names: some only initials, others full names. Perhaps change to a variant of the bib style that only includes initials (this would also save quite a bit of space)
    \item Inconsistent use of abbreviation of journal and conference names
    \item Some conferences have missing pages numbers, or incorrect pages numbers (e.g. those with pages 1--5) 
    \item Missing conference name (e.g. just ICCV in [12])
\end{itemize}
}
\rmksv{Thanks! Fixed most of these points. I am not sure about the page numbers 1--5. The bib on the IEEE website seems to say the same information as what I have. Would you be able to fix the specific onew that are wrong ?}
\rmkbw{I fixed the pp. 1--5 issue: I see why you couldn't find the problem - they were incorrect on IEEE Explore. One truly trivial remaining nitpick that certainly doesn't have to be addressed is that journal titles are abbreviated but conference names are not. If you think it's worth addressing, we could just replace all "Proceedings of the" with "Proc.".}\rmksv{Fixed}

%% file: convDictTomo.bbl
\begin{thebibliography}{10}

\bibitem{VenkatHAADF13}
S.~Venkatakrishnan, L.~Drummy, M.~Jackson, M.~De~Graef, J.~Simmons, and
  C.~Bouman,
\newblock ``A model based iterative reconstruction algorithm for high angle
  annular dark field - scanning transmission electron microscope ({HAADF-STEM})
  tomography,''
\newblock {\em IEEE Trans. on Image Processing}, vol. 22, no. 11, Nov. 2013.

\bibitem{SaBo92}
K.~Sauer and C.~Bouman,
\newblock ``{B}ayesian {E}stimation of {T}ransmission {T}omograms {U}sing
  {S}egmentation {B}ased {O}ptimization,''
\newblock {\em IEEE Trans. on Nuclear Science}, vol. 39, pp. 1144--1152, 1992.

\bibitem{zhang2014nonlocal}
H.~Zhang, J.~Ma, Y.~Liu, H.~Han, L.~Li, J.~Wang, and Z.~Liang,
\newblock ``Nonlocal means-based regularizations for statistical {CT}
  reconstruction,''
\newblock in {\em Medical Imaging 2014: Physics of Medical Imaging}, 2014, vol.
  9033, p. 903337.

\bibitem{xu2012low}
Q.~Xu, H.~Yu, X.~Mou, L.~Zhang, J.~Hsieh, and G.~Wang,
\newblock ``Low-dose {X}-ray {CT} reconstruction via dictionary learning,''
\newblock {\em IEEE Trans. on Medical Imaging}, vol. 31, no. 9, pp. 1682--1697,
  2012.

\bibitem{zhang2016gaussian}
R.~Zhang, D.~H. Ye, D.~Pal, J.-B. Thibault, K.~D. Sauer, and C.~A. Bouman,
\newblock ``A {G}aussian mixture {MRF} for model-based iterative reconstruction
  with applications to low-dose {X}-ray {CT},''
\newblock {\em IEEE Trans. on Computational Imaging}, vol. 2, no. 3, pp.
  359--374, 2016.

\bibitem{zheng2016low}
X.~Zheng, Z.~Lu, S.~Ravishankar, Y.~Long, and J.~A. Fessler,
\newblock ``Low dose {CT} image reconstruction with learned sparsifying
  transform,''
\newblock in {\em Proc. of the IEEE Image, Video, and Multidimensional Signal
  Processing Workshop (IVMSP)}, July 2016.

\bibitem{luo20162}
J.~Luo, H.~Eri, A.~Can, S.~Ramani, L.~Fu, and B.~De~Man,
\newblock ``2.5{D} dictionary learning based computed tomography
  reconstruction,''
\newblock in {\em Anomaly Detection and Imaging with X-Rays (ADIX)}, 2016, vol.
  9847, p. 98470L.

\bibitem{lewicki1999coding}
M.~S. Lewicki and T.~J. Sejnowski,
\newblock ``Coding time-varying signals using sparse, shift-invariant
  representations,''
\newblock in {\em Advances in neural information processing systems}, 1999, pp.
  730--736.

\bibitem{zeiler2010deconvolutional}
M.~D. Zeiler, D.~Krishnan, G.~W. Taylor, and R.~Fergus,
\newblock ``Deconvolutional networks,''
\newblock in {\em Proc. of the IEEE Conference on Computer Vision and Pattern
  Recognition (CVPR)}, 2010, pp. 2528--2535.

\bibitem{wohlberg2016efficient}
B.~Wohlberg,
\newblock ``Efficient algorithms for convolutional sparse representations,''
\newblock {\em IEEE Trans. on Image Processing}, vol. 25, no. 1, pp. 301--315,
  Jan. 2016.

\bibitem{bristow2013fast}
H.~Bristow, A.~Eriksson, and S.~Lucey,
\newblock ``Fast convolutional sparse coding,''
\newblock in {\em Proceedings of the IEEE Conference on Computer Vision and
  Pattern Recognition}, 2013, pp. 391--398.

\bibitem{wohlberg2014efficient}
B.~Wohlberg,
\newblock ``Efficient convolutional sparse coding,''
\newblock in {\em Proc. of IEEE International Conference on Acoustics, Speech,
  and Signal Processing (ICASSP)}, Florence, Italy, May 2014, pp. 7173--7177.

\bibitem{garcia2018convolutional}
C.~Garcia-Cardona and B.~Wohlberg,
\newblock ``Convolutional dictionary learning: A comparative review and new
  algorithms,''
\newblock {\em IEEE Trans. on Computational Imaging}, vol. 4, no. 3, pp.
  366--381, 2018.

\bibitem{papyan2017convolutional}
V.~Papyan, Y.~Romano, M.~Elad, and J.~Sulam,
\newblock ``Convolutional dictionary learning via local processing.,''
\newblock in {\em Proc. of the IEEE International Conference on Computer Vision
  (ICCV)}, 2017, pp. 5306--5314.

\bibitem{chun2018convolutional}
I.~Y. Chun and J.~A. Fessler,
\newblock ``Convolutional dictionary learning: Acceleration and convergence,''
\newblock {\em IEEE Trans. on Image Processing}, vol. 27, no. 4, pp.
  1697--1712, 2018.

\bibitem{degraux2017online}
K.~Degraux, U.~S. Kamilov, P.~T. Boufounos, and D.~Liu,
\newblock ``Online convolutional dictionary learning for multimodal imaging,''
\newblock in {\em Proc. of the IEEE International Conference on Image
  Processing (ICIP)}, 2017, pp. 1617--1621.

\bibitem{liu2018first}
J.~Liu, C.~Garcia-Cardona, B.~Wohlberg, and W.~Yin,
\newblock ``First and second order methods for online convolutional dictionary
  learning,''
\newblock {\em SIAM Journal on Imaging Sciences}, vol. 11, no. 2, pp.
  1589--1628, 2018.

\bibitem{gu2015convolutional}
S.~Gu, W.~Zuo, Q.~Xie, D.~Meng, X.~Feng, and L.~Zhang,
\newblock ``Convolutional sparse coding for image super-resolution,''
\newblock in {\em Proc. of the IEEE International Conference on Computer Vision
  (ICCV)}, 2015, pp. 1823--1831.

\bibitem{liu2016image}
Y.~Liu, X.~Chen, R.~K. Ward, and Z.~J. Wang,
\newblock ``Image fusion with convolutional sparse representation,''
\newblock {\em IEEE Signal Processing Letters}, vol. 23, no. 12, pp.
  1882--1886, 2016.

\bibitem{serrano2016convolutional}
A.~Serrano, F.~Heide, D.~Gutierrez, G.~Wetzstein, and B.~Masia,
\newblock ``Convolutional sparse coding for high dynamic range imaging,''
\newblock {\em Computer Graphics Forum}, vol. 35, no. 2, pp. 153--163, May
  2016.

\bibitem{zhang2017convolutional}
H.~Zhang and V.~M. Patel,
\newblock ``Convolutional sparse and low-rank coding-based rain streak
  removal,''
\newblock in {\em Proc. IEEE Winter Conference on Applications of Computer
  Vision (WACV)}, March 2017.

\bibitem{quan2016compressed}
T.~M. Quan and W.-K. Jeong,
\newblock ``Compressed sensing reconstruction of dynamic contrast enhanced
  {MRI} using {GPU}-accelerated convolutional sparse coding,''
\newblock in {\em Proc. of the IEEE International Symposium on Biomedical
  Imaging (ISBI)}, 2016, pp. 518--521.

\bibitem{skau2018tomographic}
E.~Skau and C.~Garcia-Cardona,
\newblock ``Tomographic reconstruction via {3D} convolutional dictionary
  learning,''
\newblock in {\em Proc. of the IEEE Image, Video, and Multidimensional Signal
  Processing Workshop (IVMSP)}, June 2018.

\bibitem{de2018tomobank}
F.~De~Carlo, D.~G{\"u}rsoy, D.~J. Ching, K.~J. Batenburg, W.~Ludwig,
  L.~Mancini, F.~Marone, R.~Mokso, D.~M. Pelt, J.~Sijbers, et~al.,
\newblock ``Tomobank: a tomographic data repository for computational {X}-ray
  science,''
\newblock {\em Measurement Science and Technology}, vol. 29, no. 3, pp. 034004,
  2018.

\bibitem{VenkatPlugPlay13}
S.~V. Venkatakrishnan, C.~A. Bouman, and B.~Wohlberg,
\newblock ``Plug-and-play priors for model based reconstruction,''
\newblock in {\em Proc. of IEEE Global Conference on Signal and Information
  Processing (GlobalSIP)}, Austin, TX, USA, Dec. 2013.

\bibitem{sreehari-2016-plug}
S.~Sreehari, S.~V. Venkatakrishnan, B.~Wohlberg, G.~T. Buzzard, L.~F. Drummy,
  J.~P. Simmons, and C.~A. Bouman,
\newblock ``Plug-and-play priors for bright field electron tomography and
  sparse interpolation,''
\newblock {\em IEEE Trans. on Computational Imaging}, vol. 2, no. 4, pp.
  408--423, Dec. 2016.

\bibitem{wohlberg2016convolutional2}
B.~Wohlberg,
\newblock ``Convolutional sparse representations as an image model for impulse
  noise restoration,''
\newblock in {\em Proc. of the IEEE Image, Video, and Multidimensional Signal
  Processing Workshop (IVMSP)}, Bordeaux, France, July 2016.

\bibitem{wohlberg2017sporco}
B.~Wohlberg,
\newblock ``{SPORCO}: {A} {P}ython package for standard and convolutional
  sparse representations,''
\newblock in {\em Proc. of the 15th Python in Science Conference}, Austin, TX,
  USA, July 2017, pp. 1--8.

\bibitem{wohlberg-2017-convolutional3}
B.~Wohlberg,
\newblock ``Convolutional sparse coding with overlapping group norms,''
\newblock Tech. {R}ep. 1708.09038, arXiv, Aug. 2017.

\bibitem{boyd2011distributed}
S.~Boyd, N.~Parikh, E.~Chu, B.~Peleato, and J.~Eckstein,
\newblock ``Distributed optimization and statistical learning via the
  alternating direction method of multipliers,''
\newblock {\em Foundations and Trends{\textregistered} in Machine learning},
  vol. 3, no. 1, pp. 1--122, 2011.

\bibitem{wohlberg2016sporco}
B.~Wohlberg,
\newblock ``{SP}arse {O}ptimization {R}esearch {CO}de ({SPORCO}),'' Software
  library available from \url{http://purl.org/brendt/software/sporco}, 2016.

\bibitem{BleichrodtAstra16}
F.~Bleichrodt, T.~van Leeuwen, W.~J. Palenstijn, W.~van Aarle, J.~Sijbers, and
  K.~J. Batenburg,
\newblock ``Easy implementation of advanced tomography algorithms using the
  {ASTRA} toolbox with {S}pot operators,''
\newblock {\em Numerical Algorithms}, vol. 71, no. 3, pp. 673--697, March 2016.

\bibitem{KimOGM15}
D.~Kim and J.~A. Fessler,
\newblock ``An optimized first-order method for image restoration,''
\newblock in {\em Proc. of the IEEE International Conference on Image
  Processing (ICIP)}, Sept. 2015, pp. 3675--3679.

\end{thebibliography}
